\documentclass[5p,preprint]{elsarticle} 
\usepackage{amsmath}
\usepackage{amssymb}
\usepackage{booktabs}
\usepackage{caption}
\usepackage{graphicx}
\usepackage[hyphens]{url}
\usepackage[colorlinks=true]{hyperref}
\usepackage[utf8]{inputenc}
\usepackage[super]{nth}
\usepackage[separate-uncertainty=true]{siunitx}
\usepackage{xurl}

\biboptions{numbers,sort&compress}

\begin{document}

\title{Correlated subgrain and particle analysis of a recovered Al-Mn alloy by directly combining EBSD and backscatter electron imaging}
\date{\today}

\author[1]{Håkon W. Ånes\corref{corr_author}}
\ead{hwaanes@gmail.com}
\author[2]{Antonius T. J. van Helvoort}
\author[1]{Knut Marthinsen}

\cortext[corr_author]{Corresponding author}
\address[1]{Department of Materials Science and Engineering, Norwegian University of Science and Technology, 7491, Trondheim, Norway}
\address[2]{Department of Physics, Norwegian University of Science and Technology, 7491, Trondheim, Norway}

\journal{Materials Characterization}


\begin{abstract}
Correlated analysis of (sub)grains and particles in alloys is important to understand transformation processes and control material properties.
A multimodal data fusion workflow directly combining subgrain data from electron backscatter diffraction (EBSD) and particle data from backscatter electron (BSE) images in the scanning electron microscope is presented.
The BSE images provide detection of particles smaller than the applied step size of EBSD down to \SI{0.03}{\micro\metre} in diameter.
The workflow is demonstrated on a cold-rolled and recovered Al-Mn alloy, where constituent particles formed during casting and dispersoids formed during subsequent heating affect recovery and recrystallization upon annealing.
The multimodal dataset enables statistical analysis including subgrains surrounding constituent particles and dispersoids' location with respect to subgrain boundaries.
Among the subgrains of recrystallization texture, Cube\{001\}$\left<100\right>$ subgrains experience an increased Smith-Zener drag from dispersoids on their boundaries compared to CubeND\{001\}$\left<310\right>$ and P\{011\}$\left<\bar{5}\bar{6}6\right>$ subgrains, with the latter experiencing the lowest drag.
Subgrains at constituent particles are observed to have a growth advantage due to a lower dislocation density and higher boundary misorientation angle.
The dispersoid size per subgrain boundary length increases as a function of misorientation angle.
The workflow should be applicable to other alloy systems where there is a need for analysis correlating grains and grain boundaries with secondary phases smaller than the applied EBSD step size but resolvable by BSE imaging.
\end{abstract}






\maketitle


\section{Introduction}

Electron backscatter diffraction (EBSD) is a technique to characterize crystallographic features in the scanning electron microscope (SEM) \cite{schwartz2009electron}.
It is routinely used to characterize (sub)grain structures and textures in metals \cite{humphreys2001grain}, which strongly affect their mechanical properties.
Secondary phases in metals, like larger constituent particles or closely spaced fine dispersoid particles, also strongly affect the mechanical properties of metals \cite{huang2018double}.
These phases are often characterized by backscatter electron (BSE) imaging in the SEM, utilizing the images' mean atomic number contrast \cite{goldstein2017scanning} which enable their separation from the matrix phase.
However, the analysis of grains and secondary phases is typically performed on separate datasets which are not directly spatially correlated.
Without this correlation, the study of the effect of individual particles on individual grains is difficult to automate and upscale to become statistically significant.
Furthermore, BSE imaging has a higher spatial resolution than EBSD in general, of a few nm compared to 20-50 nm for EBSD, and can provide information on phases smaller than those resolvable by EBSD.

Generally, in alloys containing a large number of secondary phase particles, large constituent particles ($\geq$ \SI{1}{\micro\metre}) promote recrystallization by particle stimulated nucleation (PSN) \cite{humphreys1977nucleation}, while closely spaced fine dispersoids (sub-micron) inhibit the process by pinning (sub)grain boundaries via Smith-Zener pinning \cite{doherty1997recrystallization,smith1948grains}.
Dispersoid particles precipitated concurrently with recrystallization in cold-rolled Al-Mn alloys annealed either at a low temperature or non-isothermally have a pronounced effect on their recrystallized grain structure and texture \cite{tangen2010effect,zhao2016orientation,huang2017controlling}.
Observations include sluggish recrystallization, a coarse grain structure elongated along the rolling direction, and an unusually sharp P\{011\}$\left<\bar{5}\bar{6}6\right>$ texture component and a less sharp CubeND\{001\}$\left<310\right>$ component, but still stronger compared to the Cube\{001\}$\left<100\right>$ component.
These observations are explained by a growth advantage of P subgrains during recovery and recrystallization, assumed to come about by their boundaries experiencing a reduced Smith-Zener drag than those of CubeND and Cube subgrains \cite{tangen2010effect,huang2017controlling}.
Direct statistically significant evidence of this reduced drag has not yet been reported.

The goal of this work is to establish a data fusion workflow to directly spatially correlate secondary phases, smaller than the applied EBSD step size, and orientation mapping data of (sub)grains and (sub)grain boundaries.
The workflow is demonstrated on a 95\% cold-rolled and non-isothermally annealed Al-Mn alloy to a recovered state just before the onset of recrystallization.
The resulting multimodal dataset enables correlated analysis of dispersoid and constituent particles and subgrains, and the results are presented and discussed in the context of the effect of these particles on the three recrystallization texture components P, CubeND, and Cube.


\section{Methods}

\subsection{Material and sample preparation}

The Al-Mn alloy, supplied by Hydro Aluminium, is direct-chill cast and has a composition of Al--0.53Fe--0.39Mn--0.152Si wt.\%.
It is homogenized and cold-rolled to 95\% reduction.
Huang \textit{et al.} \cite{huang2017controlling} studied the exact same alloy applying the same homogenization and deformation, and found that it was fully recrystallized after non-isothermal annealing from room temperature to \SI{350}{\celsius} at a rate of \SI{50}{\celsius\per\hour}.
Preliminary analysis shows the same results for the current study.
The material has in this state a coarse and elongated grain structure and a dominating P texture.
In this work, the material is annealed to \SI{300}{\celsius} in an air furnace and water quenched, again following the same annealing procedure as Huang \textit{et al.}.
The material has in this state recovered and is on the onset of recrystallization.

The sample surface must be sufficiently free of deformation for EBSD analysis and provide good phase contrast in BSE images to allow particle detection.
The sample surface inspected is the plane spanned by the rolling direction (RD) and normal direction (ND).
The sample embedded in epoxy is metallographically polished using diamond paste down to \SI{1}{\micro\metre} grain size, followed by vibration polishing in a \textit{Buehler VibroMet 2} using colloidal silica of \SI{0.05}{\micro\metre} grain size.
The final deformed surface layer is removed with ion polishing in a \textit{Hitachi IM-3000} using Ar$^{+}$ ions accelerated to \SI{3}{\kilo\electronvolt} with the sample tilted \SI{60}{\degree} from the horizontal and rotating.
The sample is plasma cleaned with a \textit{Fischione 1020} prior to insertion in the SEM.

\subsection{Acquisition of EBSD data and BSE images}

Three EBSD datasets are acquired with a \textit{NORDIF UF-1100} detector on a \textit{Zeiss Ultra 55} FEG SEM operated at an acceleration voltage of \SI{17}{\kilo\volt}.
The acceleration voltage is a compromise between the spatial resolution, which in general decreases with increasing voltage, and the BSE yield on the detector, which increases with increasing voltage.
The sample is tilted \SI{70}{\degree} from the horizontal towards the detector, and the working distance is about \SI{24}{\milli\metre} for all datasets.
All datasets are acquired from nominally square regions of interest (ROIs) in the middle of the rolled slab with a nominal step size $\Delta_{\mathrm{EBSD}}$ = \SI{0.1}{\micro\metre}.
The nominal area analyzed is \SI{0.021}{\milli\metre\squared}.
The detector is binned by a factor of 6 to a pattern resolution of (96 $\times$ 96) pixels of 8-bit depth.
To determine the average projection center (PC) for each dataset, five calibration patterns of (240 $\times$ 240) pixels are collected from each corner and the center of the ROI prior to acquisition.
The acquisition speed is 70 or 75 patterns per second.
The EBSD acquisition parameters are chosen to both provide a sufficient signal-to-noise ratio on the detector and a reasonable acquisition time.
Pattern analysis is detailed in Sec. \ref{sec:analysis-ebsd-patterns-orientations}.

BSE images covering each ROI are acquired after each EBSD acquisition.
The sample is reversed to a horizontal position, the working distance is reduced to \SI{5.4}{\milli\metre}, and the acceleration voltage is reduced to \SI{5}{\kilo\volt}.
Four BSE images with sufficient overlap between the images are acquired.
The acquisition time of each image is \SI{3}{\minute}.
The images are stitched into one image, henceforth called the `BSE image', using the \texttt{ImageJ} plugin \texttt{BigStitcher} \cite{horl2019bigstitcher}.
The BSE image appears free of stitching errors.
Four images are acquired instead of just one in order to use a smaller step size $\Delta_{\mathrm{BSE}}\sim$\SI{0.026}{\micro\metre}.
This is important as the Al-Mn alloy contains dispersoids smaller than $\Delta_{\mathrm{EBSD}}$.
A smaller step size could be used to detect finer dispersoids, but this would require more BSE images to be overlapped to cover the ROI, increasing the total BSE image acquisition time and the risk of stitching errors.

\subsection{Multimodal data fusion workflow}

This workflow has two goals: (1) correct distortions in the EBSD map and (2) determine whether a point in the corrected EBSD map contains a particle.
The former is accomplished by image registration, while the latter is accomplished by particle detection and data fusion accounting for the different step sizes in the BSE image and EBSD map.
The workflow is demonstrated on one of the three datasets analyzed.
Figures for the two remaining datasets are included in the supplementary material.

\subsubsection{Image registration}

The goal of image registration \cite{goshtasby2012image} is to obtain a transformation function which can interpolate between coordinates in a `reference' image and a `sensed' image, here the BSE image and EBSD map, respectively.
Ideally, information from the BSE image could be inserted directly into the EBSD map as long as the difference in spatial resolution is handled by binning.
However, the EBSD map is generally distorted \cite{nolze2007image} compared to the BSE image, mostly due to the high sample tilt and longer acquisition time.
The thin plate spline (TPS) transformation function \cite{bookstein1989principal,goshtasby2012image} previously used by Zhang \textit{et al.} \cite{zhang2014method} to correct distortions in EBSD maps is used for image registration in this work.

\begin{figure*}[htb]
  \centering
  \includegraphics[width=\textwidth]{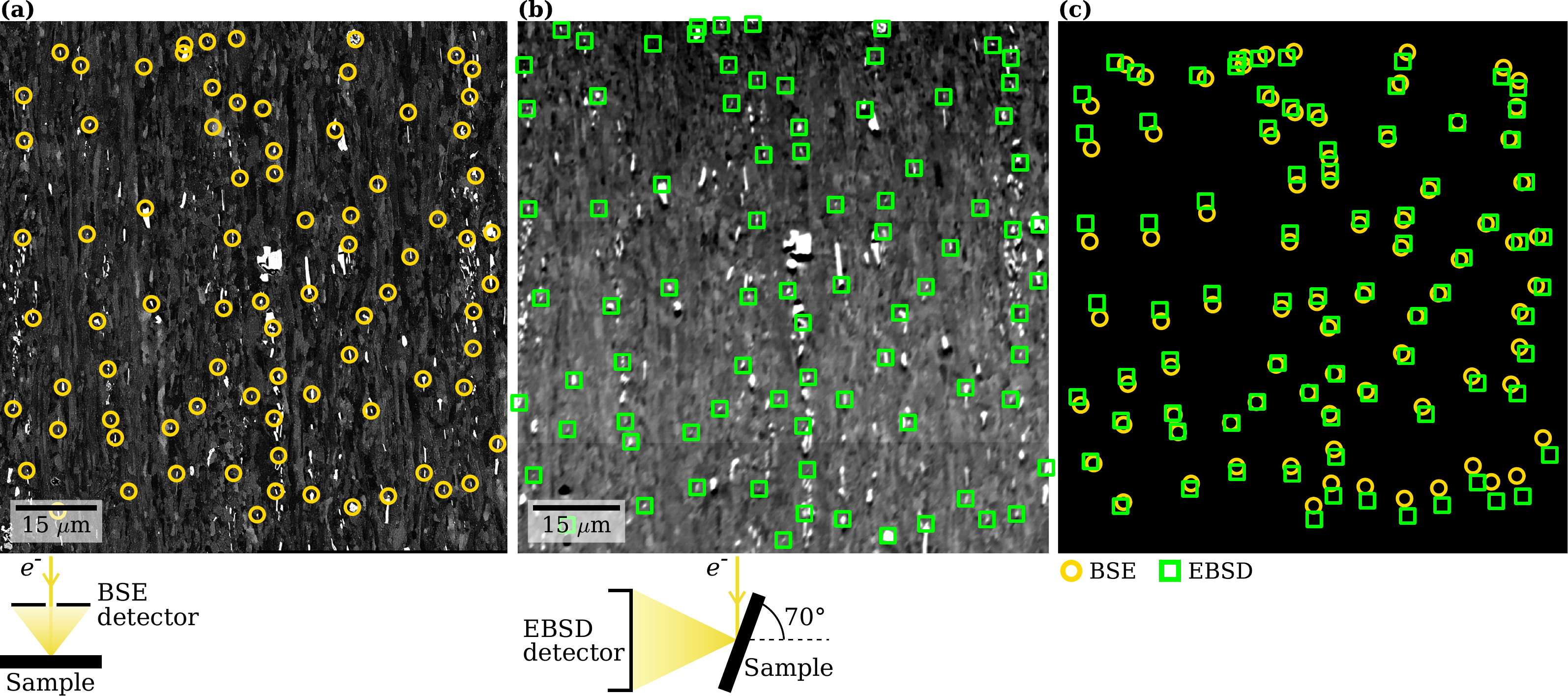}
  \caption{
    Control points used in image registration.
    82 manually selected control points in the same ROI in (a) the BSE image, (b) the EBSD intensity map, and (c) their relative positions after scaling and translating the EBSD control point coordinates.
    Positions of the sample and detectors relative to the electron beam during BSE image and EBSD map acquisition are shown (not to scale) below (a) and (b), respectively.
  }
  \label{fig:control-points}
\end{figure*}

The general starting point of image registration is the reference and sensed images of the same ROI showing features recognizable in both images \cite{goshtasby2012image}.
The images used here are the BSE image, which shows mostly mean atomic number contrast, and an EBSD intensity map obtained by summing the intensities in each raw EBSD pattern.
A part of the full BSE image with the ROI and the EBSD intensity map, from the first of the three datasets analyzed, are given in Fig. \ref{fig:control-points} (a) and (b), respectively.
The contrast in the EBSD intensity map is stretched by clipping 1\% of the lowest and highest intensities.
Also presented are the sample and detector positions relative to the electron beam during BSE image and EBSD data acquisition.

The TPS transformation function requires corresponding control points (CPs) in the BSE image and EBSD map.
Larger particles quite evenly distributed on the sample surface are recognizable in both images in Fig. \ref{fig:control-points} (a, b), and so 82 CPs are manually selected mostly from particle locations, plotted in (a, b).
Not all of the recognizable particles are selected as CPs; the selection of CPs is discussed in Sec. \ref{sec:discussion-workflow}.
Relative positions of both sets of CPs are plotted in Fig. \ref{fig:control-points} (c), where the CP coordinates in the BSE image are first downscaled to the same scale as the EBSD map by dividing by $\Delta_{\mathrm{EBSD}} / \Delta_{\mathrm{BSE}}$, and then translated by subtracting the average shift between the CP coordinates in the BSE image and the EBSD map.
If the coordinates of $n$ CPs are denoted $(x_i, y_i)$ in the BSE image and $(X_i, Y_i)$ in the EBSD map, the TPS transformation function interpolating between the two sets of CPs is

\begin{equation}
  (X, Y) = a_1 + a_x x + a_y y + \sum_{i=1}^nw_iU(\left|(x_i,y_i) - (x,y)\right|),
  \label{eq:thin-plate-spline}
\end{equation}

\noindent where the kernel function is defined as $U(R) = R^2\log R^2$.
The function represents a plate of infinite extent deforming under point loads at the CPs \cite{goshtasby2012image}. The first three terms, an affine transformation, define the plate best matching all the CPs, while the last term, a weighted sum of radially symmetric basis functions, represent bending forces provided by the $n$ CPs.
The unknown coefficients $a_1$, $a_x$, $a_y$, and $w_i$ are determined based on the coordinates of the two sets of CPs, and the procedure is detailed by Zhang \textit{et al.} \cite{zhang2014method}.
We determine them using the TPS transformation function implemented in the \texttt{Python} package \texttt{morphops} v0.1.13 \cite{morphops}, which is based on the work by Bookstein \cite{bookstein1989principal}.
The inverse mapping from the downscaled BSE image coordinates $(x_i, y_i)$ to the EBSD map coordinates $(X_i, Y_i)$ is then found by inserting $(x_i, y_i)$ into Eq. \eqref{eq:thin-plate-spline} together with the determined coefficients.
This mapping allows points in the EBSD map to be re-mapped to the scaled BSE image coordinates by rounding to the closest map coordinate, which requires no interpolation of information in the original EBSD map points, like orientations.

\begin{figure*}[htb]
  \centering
  \includegraphics[width=\textwidth]{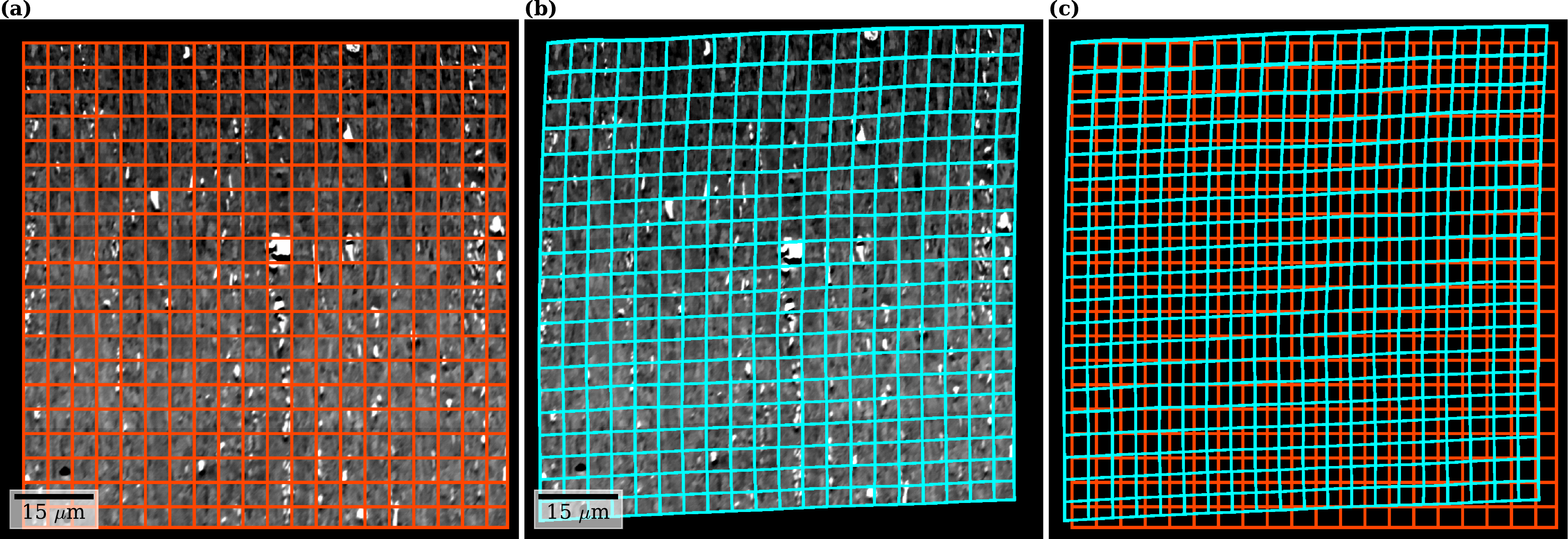}
  \caption{
  EBSD intensity map before and after correction using the TPS transformation function and the CPs in Fig. \ref{fig:control-points}.
  The EBSD intensity map with a grid attached (a) before correction and (b) after correction, and (c) the grid before and after correction.
  }
  \label{fig:corrected-grid}
\end{figure*}

The EBSD intensity map before and after correction is shown in Fig. \ref{fig:corrected-grid} (a) and (b), respectively.
A rectangular grid with a nominal spacing of \SI{5.1}{\micro\metre} is attached to the EBSD intensity map before correction in order to highlight the distortions, and the grids before and after correction are presented in (c).
The nominal area analyzed is \SI{8446}{\micro\metre\squared}, but the actual area scanned is \SI{8123}{\micro\metre\squared}, which is a 4\% reduction.
The actual area scanned was 7\% greater compared to the nominal area for the other two datasets, demonstrating that the distortions are not constant between scans.

\subsubsection{Detection of particles and data fusion}

Knowing the coordinates of each EBSD map point in the BSE image, it is possible to insert information from the latter in the EBSD map.
To utilize the higher fidelity of the BSE image, particles are detected in the BSE image in its original resolution, also outside the EBSD map ROI to improve the statistical analysis of particle size and volume fraction.
This produces a binary particle map, and for the ROI in this particle map to fit into the EBSD map, it must be binned.
If the binned particle map is to remain binary, the binning factor must be an integer.
The different spatial resolutions $\Delta_{\mathrm{EBSD}}$ and $\Delta_{\mathrm{BSE}}$ do not allow this, so prior to particle detection, the BSE image is upscaled to allow integer binning.
The integer binning factor is rounded to the nearest upper integer $x_{\mathrm{bin}} = \lceil\Delta_{\mathrm{EBSD}} / \Delta_{\mathrm{BSE}}\rceil$, and the upscaling factor is $x_{\mathrm{bin}} \cdot (\Delta_{\mathrm{BSE}} / \Delta_{\mathrm{EBSD}})$.

Particles are detected in the upscaled BSE image using region-based segmentation as implemented in the \texttt{Python} package \texttt{scikit-image} v0.18 \cite{walt2014scikit-image}.
Particles are detected in the BSE images from all three datasets following six steps with all parameters fixed:
(1) Remove long-range intensity variations by subtracting a Gaussian blurred version of the image, and normalize the image intensities;
(2) Generate an elevation map using the amplitude of the intensity gradient obtained by correlating the BSE image with a Sobel filter;
(3) Determine markers of points that can unambiguously be labelled as particle or background (subgrains) in the image by applying a threshold found from the intensity distribution;
(4) Segment particles with the watershed transform using the elevation map and markers;
(5) Remove holes within segmented particles, as they are assumed to contain no holes;
(6) Finally, remove incorrectly detected particles by assuming that correctly detected particles have (i) sufficiently high roundness, defined as the ratio between the perimeter of the convex hull and the actualt perimeter, (ii) sufficiently high solidity, defined as the ratio between the area of the convex hull and the actual area, and (iii) sufficiently high mean BSE image intensity compared to the background.
Based on these assumptions, particles which satisfy \textit{any} of the following three criteria are kept: (i) roundness greater than the \nth{25} percentile, (ii) solidity greater than the \nth{25} percentile, and (iii) mean BSE image intensity greater than the \nth{55} percentile.
By identifying some correctly and incorrectly detected particles, the percentile thresholds are found manually in a search so that as many of the former and as few of the latter are kept.

\begin{figure*}[htb]
  \centering
  \includegraphics[width=\textwidth]{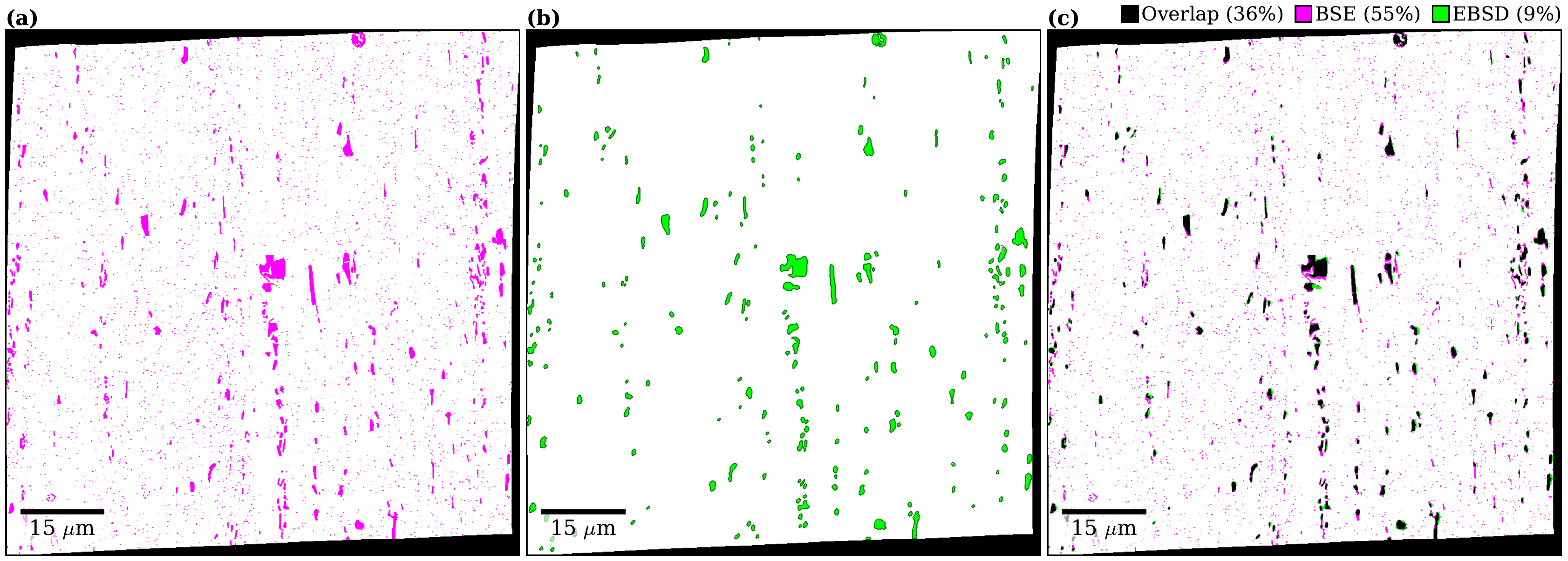}
  \caption{
  Particles detected with region-based segmentation in the ROI defined by the corrected EBSD map.
  Particles detected in (a) the BSE image and (b) the corrected EBSD intensity map, and (c) the two particle sets overlapped, with overlapping regions shown in black.
  }
  \label{fig:overlapping-particles}
\end{figure*}

To get a qualitative assessment of how good the image registration and particle detection is, particles are detected in the EBSD intensity map in Fig. \ref{fig:control-points} (a) following the steps (1-5) above.
The 4972 particles detected in the BSE image and the 217 particles detected in the EBSD intensity map are shown in Fig. \ref{fig:overlapping-particles} (a, b), respectively, while the particles are given in (c), with overlapping particles colored black.
Many of the larger particles detected in both images overlap, accounting for 36\% of colored pixels in (c).
This indicates a reasonably successful image registration and that particles are correctly detected, assuming a similar mean atomic number contrast in both images.
There are thousands of small particles and some larger ones only detected in the BSE image, visible as pink in (c), accounting for 55\% of colored pixels.
The 9\% of colored pixels only detected in the EBSD intensity map, visible as green in (c), are mostly found in pixels surrounding particles detected in both the BSE image and the EBSD intensity map.
This can be explained by the fact that the detected signal on the EBSD detector originates from a larger interaction volume due to the higher acceleration voltage and the \SI{70}{\degree} sample tilt during EBSD data acquisition.

Before inserting the upscaled binary particle map into the EBSD map, the particle map is binned by the integer binning factor $x_{\mathrm{bin}}$, meaning blocks of ($x_{\mathrm{bin}} \times x_{\mathrm{bin}}$) pixels are binned into one pixel.
With $x_{\mathrm{bin}} = 4$, particles constituting one pixel and 16 pixels in the unbinned particle map might be given the same `size' of one pixel in the binned particle map, which will distort the analysis of particle size from the binned particle map.
To account for this, a binned map containing the number of particle pixels within each (4 $\times$ 4) block is included alongside the binned binary particle map in the EBSD data and used when calculating the particle size.
Another caveat is that if parts of two neighboring particles are binned into the same point, the number of particles in that point is two.
This fact is neglected when binning in this work, as it is only the case for fewer than 80 of more than two million points in all three datasets.
In summary, the final result of the multimodal data fusion workflow is the corrected EBSD map containing the binned binary particle map and the map of the number of particle pixels in each map point.

\subsection{Analysis of EBSD patterns and orientations}
\label{sec:analysis-ebsd-patterns-orientations}

The crystal orientations from each EBSD pattern in the original EBSD maps are determined by searching for the best match in a discrete dictionary of simulated patterns by dictionary indexing (DI) \cite{chen2015dictionary}, as implemented in the \texttt{Python} package \texttt{kikuchipy} v0.5 \cite{kikuchipy0.5,pena2017electron}.
The basis for DI is a dynamically simulated master pattern, simulated for Al (space group $Fm\bar{3}m$) with the \texttt{EMsoft} v4.3 \cite{callahan2013dynamical} \texttt{Fortran} suite of programs.
Simulated patterns in the dictionary are projected from the master pattern onto the EBSD detector using an average PC per dataset, found from the Al calibration patterns using Hough indexing in the \texttt{Python} package \texttt{PyEBSDIndex} \cite{pyebsdindex}.
The orientations of the simulated patterns are sampled using cubochoric sampling \cite{singh2016orientation} as implemented in the \texttt{Python} package \texttt{orix} v0.7 \cite{orix,johnstone2020density}.
The orientations are uniformly distributed in the fundamental zone with an average misorientation angle of \SI{1.4}{\degree}, resulting in a dictionary of about 300 000 simulated patterns.
The signal-to-noise ratio of the raw experimental patterns is increased prior to indexing by subtracting a static background, followed by subtraction of a dynamic (per pattern) background, and finally averaging each pattern with its eight nearest neighbors using a Gaussian kernel with a standard deviation of 1, as implemented in \texttt{kikuchipy} v0.5.
Every experimental pattern is compared to the dictionary via the normalized cross correlation (NCC) coefficient $r$ \cite{gonzalez2017digital}, and the orientation of the simulated pattern with the highest $r$ is chosen as the solution per pattern.
These discrete orientations are refined by letting them vary while keeping the PC fixed, and optimizing $r$.
The Nelder-Mead optimization algorithm as implemented in the \texttt{Python} package \texttt{SciPy} v1.7 \cite{virtanen2020scipy} is used.

Orientation analysis is done with the \texttt{MATLAB} toolbox \texttt{MTEX} v5.8 \cite{bachmann2011grain} on the corrected EBSD map containing the particle map.
No averaging or clean-up of orientations is applied.
The particle map is used to create a `dual phase' dataset to distinguish between points with Al and points with particles.
Subgrains are reconstructed with a misorientation angle threshold of \SI{1}{\degree}, with subgrains smaller than five points assigned to neighboring subgrains.
Grain boundaries are smoothed in five iterations, resulting in a reduced total boundary length and allowing boundary segments to have coordinates independent of the rectangular grid of the corrected EBSD map.
This results in two `grain' populations of Al subgrains and particles.
Typical rolling and recrystallization texture components in Al \cite{humphreys2017recrystallization} considered in this work are listed in Table \ref{tab:ideal-orientations}.
The subgrains are classified into these texture components based on the lowest misorientation angle between the subgrain mean orientation and the components' ideal orientations within a threshold of \SI{15}{\degree}.
Subgrains outside the threshold are considered random and termed `other' in the remainder of the text.
An orthorhombic sample symmetry valid for cold-rolled samples is assumed when assigning texture components.
Grain sizes $D$ and particle sizes $d$ are given as equivalent circular diameters $0.816 \cdot 2 \cdot \sqrt{A/\pi}$, where $A$ is the grain or particle area and the prefactor is a result of stereological considerations \cite{humphreys2017recrystallization}.

\begin{table}[htb]
  \centering
  \caption{Typical texture components in cold-rolled Al \cite{humphreys2017recrystallization} represented by Miller indices \{hkl\}$\left<uvw\right>$. In general, Br, Cu and S are deformation textures and the remaining are recrystallization textures.}
  \begin{tabular}{l c}
    \toprule
    Name & \{hkl\}$\left<uvw\right>$\\
    \midrule
    Br & \{011\}$\left<2\bar{1}1\right>$\\
    Cu & \{112\}$\left<11\bar{1}\right>$\\
    S & \{123\}$\left<634\right>$\\
    Cube & \{001\}$\left<100\right>$\\
    CubeND  & \{001\}$\left<310\right>$\\
    Goss & \{011\}$\left<100\right>$\\
    P & \{011\}$\left<\bar{5}\bar{6}6\right>$\\
    \bottomrule
  \end{tabular}
  \label{tab:ideal-orientations}
\end{table}


\section{Results}

\subsection{Multimodal dataset of subgrains and particles}

\begin{figure*}[htb]
  \centering
  \includegraphics[width=\textwidth]{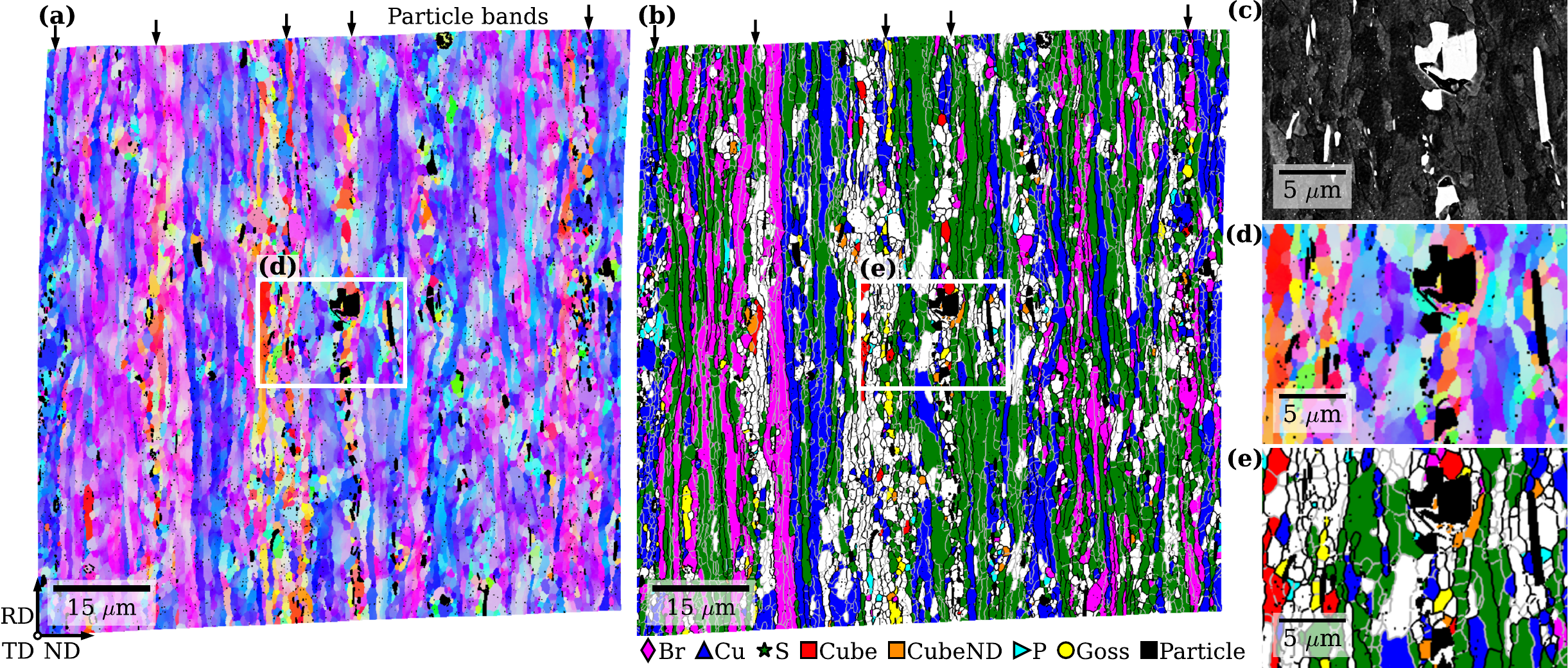}
  \caption{
  Multimodal dataset of the Al microstructure with subgrains and particles.
  (a) Al orientation map colored according to the direct lattice vector parallel to RD (indicated bottom left), and (b) subgrains colored by texture component according to the legend below.
  Subgrains not assigned a texture component are white.
  The highlighted region is magnified in (c) the BSE image and the (d) orientation and (e) grain maps.
  Particles are colored black in (a, b, d, e), while LAGB and HAGB are respectively colored gray and black in (b, e).
  Arrows in (a, b) highlight bands of constituent particles.}
  \label{fig:grains-particles}
\end{figure*}

The result of the multimodal data fusion workflow, the Al-Mn microstructure represented in terms of Al subgrains and particles, is shown in Fig. \ref{fig:grains-particles} for the first of the three datasets.
The Al orientation map colored according to the direct lattice vector parallel to RD is presented in (a, d).
Subgrains colored by texture component are shown in (b, e), with low–angle grain boundaries (LAGB) with boundary misorientation angles $\omega \in$ [\SI{1}{\degree}, \SI{15}{\degree}) and high–angle grain boundaries (HAGB) with $\omega \geq$ \SI{15}{\degree} colored gray and black, respectively.
Particles are colored black in (a, b, d, e).
Bands of constituent particles along RD are indicated with arrows in (a, b), where most of the subgrains of random orientations or recrystallization textures are located.
A magnified view of the microstructure is shown in the BSE image and orientation and grain maps in (c-e), respectively, illustrating the successful detection and localization of particles in the subgrain structure.
The microstructure is recovered, and no obvious recrystallization nuclei are observed in the three datasets studied.

\begin{figure}[htb]
  \centering
  \includegraphics[width=\columnwidth]{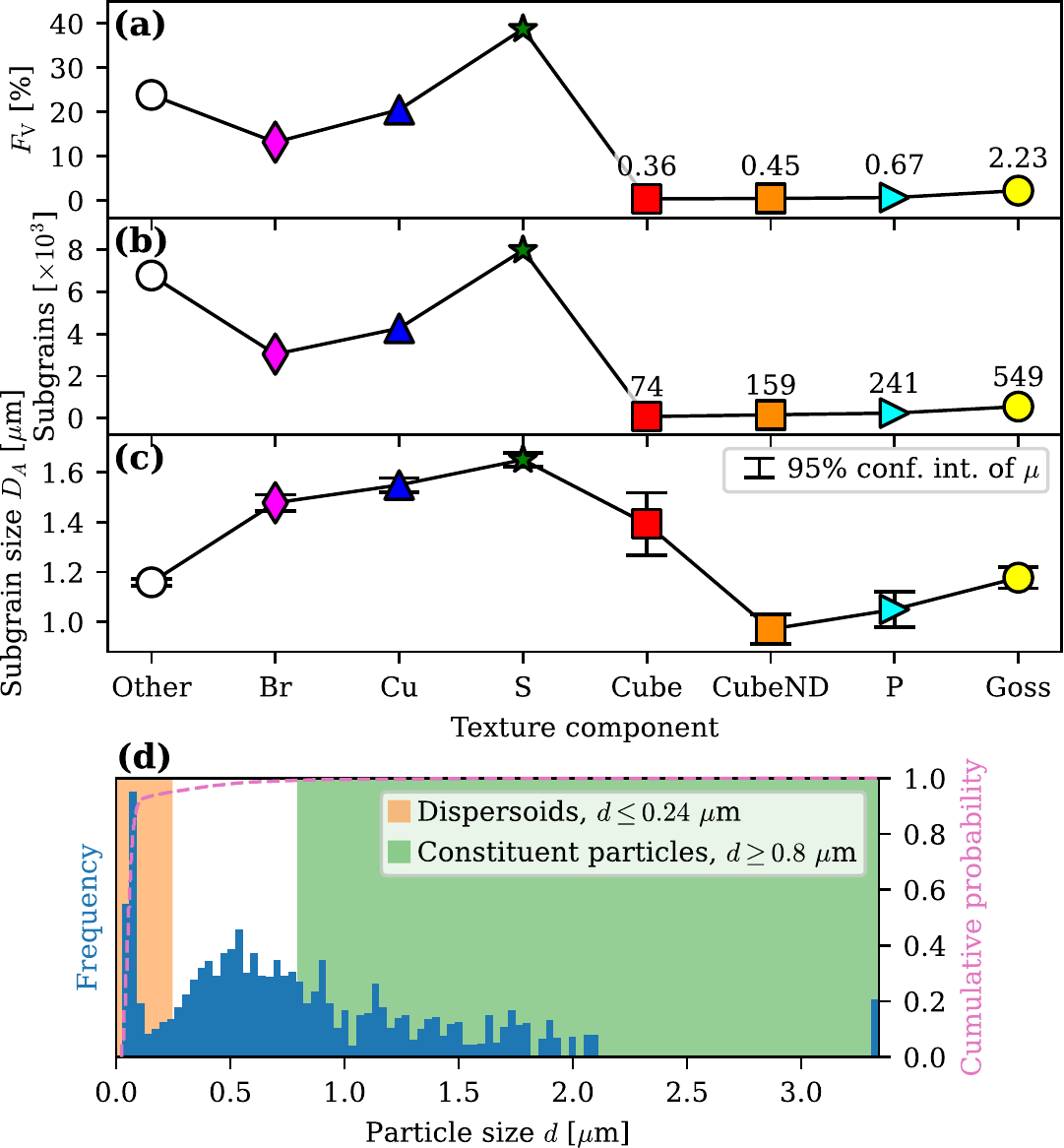}
  \caption{
  Subgrain and particle statistics from all three datasets combined.
  (a) Volume fraction $F_V$, (b) number of subgrains, and (c) area weighted grain size $D_A$, per texture component.
  (d) Area weighted histogram and cumulative probability distribution of particle sizes $d$, with size regions classified as dispersoids and constituent particles indicated.
  }
  \label{fig:grain-particle-statistics}
\end{figure}

The combined subgrain and particle statistics from all three datasets are given in Fig. \ref{fig:grain-particle-statistics}.
The volume fraction $F_{\mathrm{V}}$ of subgrains of rolling texture is 72\% as shown in (a), typical of an Al alloy deformed by cold-rolling but not yet recrystallized.
15282 subgrains make up this volume fraction given in (b), with a mean area weighted grain size of $D_{\mathrm{A,\:roll.}}$ = \SI{1.59(2)}{\micro\metre} as shown in (c).
More interesting when studying recrystallization are the 782 subgrains of recrystallization texture constituting a volume fraction of 3\%, with a mean grain size of $D_{\mathrm{A,\:rex.}}$ = \SI{1.15(3)}{\micro\metre}.
The remaining grain volume fraction are 6777 subgrains of other orientations with an average grain size of $D_{\mathrm{A,\:other}}$ = \SI{1.16(1)}{\micro\metre}.
The uncertainties are the 95\% confidence interval of the mean, and this is used throughout the text when uncertainties are given.

A histogram of the area weighted particle size $d$ of about 90000 particles detected in the BSE images from all three datasets, including outside the EBSD ROIs, is shown in Fig. \ref{fig:grain-particle-statistics} (d).
About 11000 of the particles are inside the EBSD ROIs.
The cumulative probability distribution of the particle sizes $d$ plotted in (d) shows that about 91\% of particles are smaller than \SI{0.1}{\micro\metre} in diameter, which constitutes 1 point in the EBSD maps and 17 pixels in the BSE images.
Dispersoids are here defined as those smaller than \SI{0.24}{\micro\metre}, seen in Fig. \ref{fig:grain-particle-statistics} (d).
This threshold is chosen based on the somewhat bimodal distribution of $d$.
The critical diameter for particle stimulated nucleation (PSN) after rolling to 95\% is $\sim$\SI{1}{\micro\metre} \cite{humphreys1977nucleation}, and \SI{0.8}{\micro\metre} is used here as the lower threshold for particles defined as constituent particles, which are considered potential nucleation sites for recrystallized grains.
Particle parameters for all detected particles, constituent particles, and dispersoids are given in Table \ref{tab:particle-parameters}, assuming particles to be spherical and randomly distributed.

\begin{table}[htb]
  \centering
  \caption{
  Particle parameters: volume fraction $f_{\mathrm{V}}$, area weighted average particle size $d_{\mathrm{A}}$, and interparticle spacing $N_{\mathrm{s}}^{-0.5}$.
  }
  \begin{tabular}{l c c c}
    \toprule
    Population & $f_{\mathrm{V}}$ [\%] & $d_{\mathrm{A}}$ [\si{\micro\metre}] & $N_{\mathrm{s}}^{-0.5}$ [\si{\micro\metre}] \\
    \midrule
    All & 1.84 & \SI[multi-part-units=single]{0.802(1)}{} & 0.306 \\
    Constituent & 0.73 & \SI[multi-part-units=single]{1.400(79)}{} & 10.852 \\
    Dispersoids & 0.31 & \SI[multi-part-units=single]{0.089(1)}{} & 0.723 \\
    \bottomrule
  \end{tabular}
  \label{tab:particle-parameters}
\end{table}

\subsection{Dispersoids at subgrain boundaries}

\begin{figure*}[htb]
  \centering
  \includegraphics[width=\textwidth]{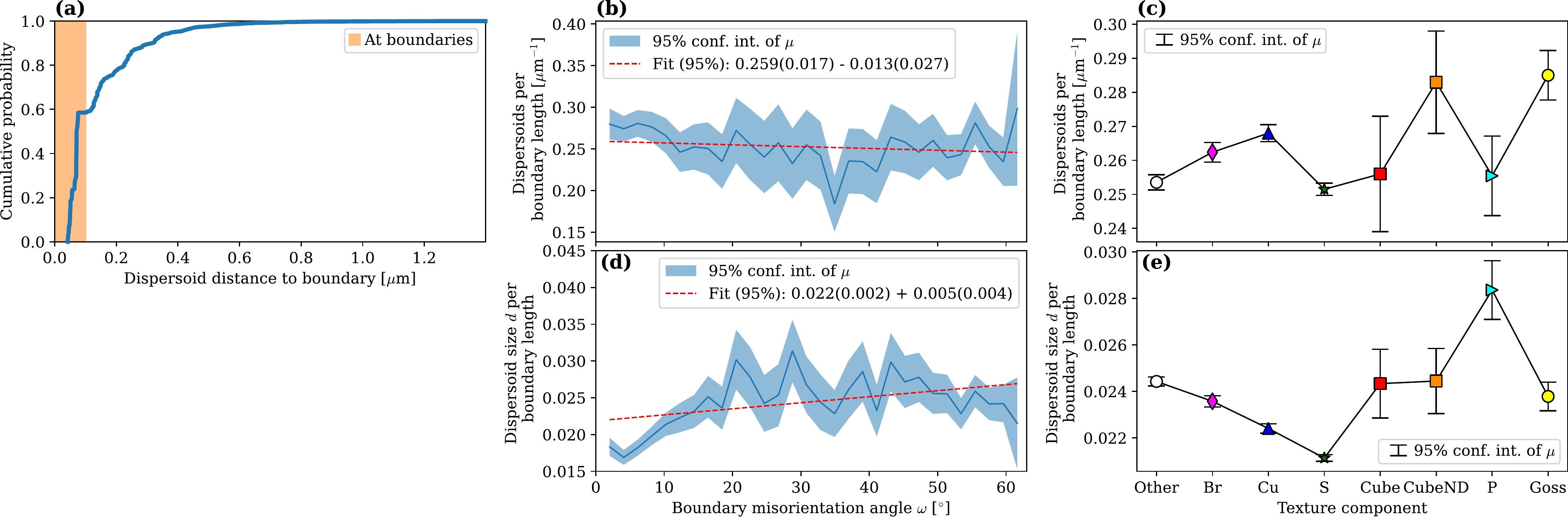}
  \caption{
  Dispersoids at subgrain boundaries.
  (a) Cumulative probability distribution of dispersoids' distance to the nearest subgrain boundary.
  Dispersoids with distances below \SI{0.1}{\micro\metre} are considered to be at boundaries.
  (b) Dispersoids per boundary length as a function of boundary misorientation angle.
  (c) Dispersoids per boundary length per texture component.
  (d) Dispersoid size $d$ per boundary length as a function of boundary misorientation angle.
  (e) Dispersoid size $d$ per boundary length per texture component.
  Linear fits are shown in (b, d) while 95\% confidence intervals are shown in (b-e).
  }
  \label{fig:dispersoids-at-boundaries}
\end{figure*}

By computing the Euclidean distance from a particle to every subgrain boundary segment within a square of (2 $\times$ 2) \si{\micro\metre} centered on the particle, the minimum distance from each particle to a segment and the number of particles within a certain distance from a segment can be determined.
The cumulative probability of dispersoids' distance to the closest boundary is given in Fig. \ref{fig:dispersoids-at-boundaries} (a): 60\% of dispersoids are within \SI{0.1}{\micro\metre} of a boundary, the same distance as $\Delta_{\mathrm{EBSD}}$, and these dispersoids are considered to be at boundaries.
The remaining dispersoids are within subgrains.
The farthest a dispersoid is from a boundary is about \SI{1.3}{\micro\metre}, which is reasonable considering that the average subgrain size $D_{\mathrm{A}}$ is of the same magnitude.
Two parameters are considered based on the dispersoids at boundaries: The number of dispersoids per boundary length, given as dispersoids \si{\per\micro\metre}, and the dispersoid size $d$ per boundary length, which is dimensionless.

Classifying boundary segments by their misorientation angle $\omega$, the number of dispersoids per boundary length per $\omega$ is plotted in Fig. \ref{fig:dispersoids-at-boundaries} (b).
There is on average 0.26 dispersoids \si{\per\micro\metre}, i.e. about 1 dispersoid every \SI{4}{\micro\metre} of boundary, and the amount stays constant with increasing $\omega$.
No correlation is observed when inspecting the misorientation axis either.
The dispersoid size $d$ per boundary length as a function of $\omega$ is plotted in (d).
There is on average about one dispersoid of size \SI{0.022}{\micro\metre} per \SI{1}{\micro\metre} of boundary, and here the amount increases slightly with increasing $\omega$.

Classifying boundary segments by the texture component on either side of the boundary, the number of dispersoids per boundary length per texture component is plotted in Fig. \ref{fig:dispersoids-at-boundaries} (c): There is no significant difference for the various subgrain boundaries, apart from slightly more dispersoids on CubeND and Goss subgrain boundaries.
Looking at the dispersoid size $d$ per boundary length as a function of texture component in (e), P subgrains stand out among the subgrains of recrystallization textures as having significantly larger dispersoids on their boundaries.

When considering only HAGBs, an insignificant difference in the number of dispersoids on HAGBs of subgrains at constituent particles and subgrains elsewhere is observed: the former boundaries have on average \\\SI{0.239(60)} dispersoids \si{\per\micro\metre} while the latter boundaries have on average \SI{0.248(7)} dispersoids \si{\per\micro\metre}.
The dispersoid size $d$ per boundary length is slightly lower but still insignificant, with \SI{0.022(6)} for HAGBs at constituent particles and \SI{0.026(1)} for HAGBs elsewhere.

P, CubeND and Cube subgrains all have orientation relationships to the deformed matrix close to the special misorientation of \SI{38}{\degree}$\left<111\right>$ \cite{engler1996nucleation}, or $\Sigma7$ in coincidence site lattice terminology, which represents a boundary with a reduced energy and a higher migration rate than general HAGBs \cite{humphreys2017recrystallization}.
A significant difference in the number of dispersoids per boundary length is observed between boundaries of $\omega \leq$ \SI{15}{\degree} to $\Sigma7$ and other boundaries: the former boundaries have on average \SI{0.233(17)}dispersoids \si{\per\micro\metre} while the latter boundaries have on average \SI{0.263(5)} \\ dispersoids \si{\per\micro\metre}.
Furthermore, the dispersoid size $d$ per boundary length is higher for $\Sigma7$ boundaries than other boundaries, with \SI{0.026(2)} compared \SI{0.022(1)}.
13-18\% of boundaries of subgrains of recrystallization textures are of $\Sigma7$-type, while less than 9\% of boundaries of subgrains of rolling textures and other orientations are of this type.

\subsection{Subgrains at constituent particles}

\begin{figure*}[htb]
  \centering
  \includegraphics[width=\textwidth]{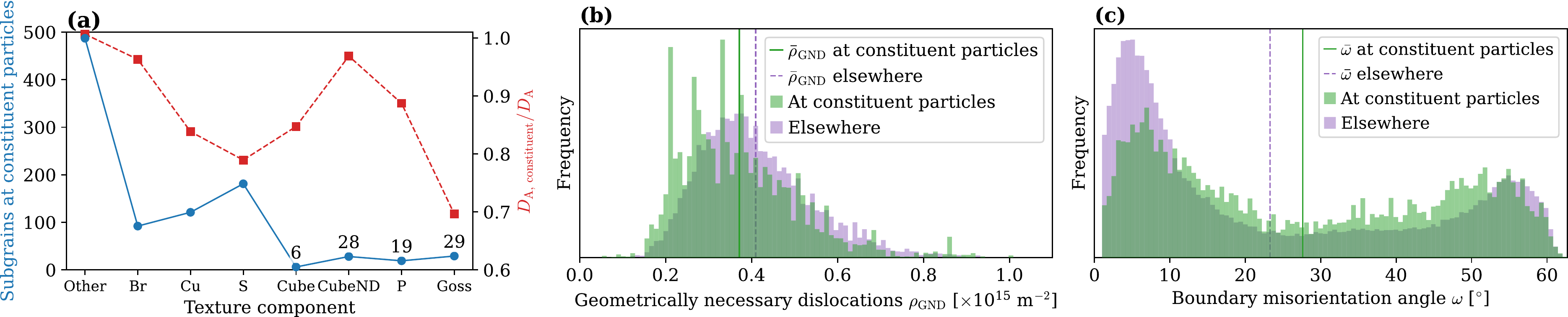}
  \caption{
  Subgrains at constituent particles.
  (a) Number of subgrains at constituent particles (left), and the ratio of mean area weighted size of subgrains at constituent particles $D_{\mathrm{A,\:constituent}}$ over all grain sizes $D_{\mathrm{A}}$ (right), per texture component.
  (b) Normalized and area weighted histograms of $\rho_{\mathrm{GND}}$ in subgrains at constituent particles and subgrains elsewhere, with the means.
  (c) Normalized histograms of the boundary misorientation angle $\omega$ of subgrains at constituent particles and subgrains elsewhere, with the means.
  }
  \label{fig:subgrains-at-constituent-particles}
\end{figure*}

Analysis of subgrains at constituent particles is presented in Fig. \ref{fig:subgrains-at-constituent-particles}.
As shown in (a), most of these subgrains are randomly oriented, which is generally found for subgrains in deformation zones at constituent particles in heavily deformed polycrystals \cite{humphreys2017recrystallization}.
This is also evident from Fig. \ref{fig:grains-particles} (b) by observing that randomly oriented subgrains are mostly found in particle bands \cite{zhang2012three}.
Subgrains of recrystallization textures at constituent particles are mostly CubeND, Goss, and P, with fewest Cube subgrains.
Examples of subgrains of recrystallization textures at constituent particles are seen in Fig. \ref{fig:grains-particles} (d, e).
Fig. \ref{fig:subgrains-at-constituent-particles} (a) shows that the sizes of subgrains per texture component at constituent particles are in general smaller than subgrains found elsewhere in the microstructure, in accordance with previous comparisons of size distributions of subgrains close to and far away from constituent particles \cite{humphreys1977nucleation}.
The amount of plastic strain within a grain can be approximated by the geometrically necessary dislocation density $\rho_{\mathrm{GND}}$, calculated from orientation differences in neighboring map points using Pantleon's approach \cite{pantleon2008resolving} implemented in \texttt{MTEX}.
As was done by Pantleon, the \{111\}$\left<110\right>$ slip system is assumed active with 12 edge dislocations with line energies of 1 and 6 screw dislocations with line energies of 1 - $\nu$, where $\nu$ = 0.347 is the Poisson ratio for Al.
Distributions of $\rho_{\mathrm{GND}}$ in subgrains at constituent particles and elsewhere are presented in Fig. \ref{fig:subgrains-at-constituent-particles} (b).
Distributions of the boundary misorientation angle $\omega$ for the same two groups of subgrains are given in Fig. \ref{fig:subgrains-at-constituent-particles} (c).
Subgrains at constituent particles have lower dislocation densities and higher $\omega$ compared to subgrains elsewhere.
This indicates that the former subgrains have recovered more than the latter subgrains, and that they have favorable conditions for becoming recrystallization nuclei, i.e. they have a growth advantage.
Two other commonly used subgrain orientation parameters, grain orientation spread and grain average misorientation, neither shown here, give a similar picture to that of $\rho_{\mathrm{GND}}$.

\section{Discussion}

\subsection{Evaluation of the multimodal data fusion workflow}
\label{sec:discussion-workflow}

The presented multimodal data fusion workflow accomplishes two tasks: first, it corrects distortions in the EBSD map \cite{nolze2007image}, and secondly, it fuses data from the particle map of higher fidelity with the corrected EBSD map.
There are numerous available solutions to the first task \cite{zhang2014method,charpagne2019accurate,tong2020trueebsd}, with various degrees of complexity.
To accomplish the first task, the present workflow requires a minimum of two images showing shared features, but which can have totally different contrast and intensity, manual selection of a sufficient number of CPs evenly distributed in the images, and a single transformation function without tuning parameters to correct distortions in the EBSD map based on the CPs.
The most critical step in the workflow is the selection of CPs.

Zhang \textit{et al.} \cite{zhang2014method} studied the effect of the number of CPs used in the TPS transformation function to correct an EBSD map, similar to the maps in this work, with $\Delta_{\mathrm{EBSD}}$ = \SI{0.2}{\micro\metre} and ROI area (120 $\times$ 120) \si{\micro\metre\squared}.
They found that correction of the EBSD map improved substantially when using 54 CPs compared to 36, and improved minimally when using 60 CPs compared to 54.
The improvement was quantified in two ways: by comparing the outline of the corrected EBSD maps to the same outline visible as a contaminated area in a BSE image obtained with no sample tilt taken as the ground truth, and by inspecting non-overlapping map points in overlapped Kikuchi band contrast maps from the corrected EBSD maps obtained when using 36 and 54 CPs to the same map when using 60 CPs, taken as the ground truth.
The surface contamination resulting from the EBSD acquisition is not visible in the BSE images in this work.
Since the end goal of image registration here is data fusion, maximizing the similarity between the reference and sensed images leads to a more correct multimodal dataset.
A similar reasoning was used by Nguyen and Rowenhorst \cite{nguyen2021alignment} in image registration and data fusion of pores from BSE images with EBSD data in an additive manufactured steel sample, when a control point-free transformation function was found iteratively by optimizing the similarity of binary maps of pores detected in the BSE images and EBSD maps.
By assuming that the BSE image and EBSD intensity map in Fig. \ref{fig:control-points} (a, b) have similar contrast, and by rescaling the corrected EBSD map to the same resolution as the BSE image and only comparing the ROI, a measure of their similarity is obtained via the NCC coefficient $r$.
The similarity between the corrected EBSD intensity map in Fig. \ref{fig:corrected-grid} (b) and the BSE image in Fig. \ref{fig:control-points} (a) when using all the 82 CPs is $r_{82}$ = 0.3625.
By using 77 CPs, chosen at random 30 times, a slightly worse $r_{77} = 0.3622 \pm 0.0003$ is obtained, and by repeating this calculation but instead using 72 CPs, $r_{72} = 0.3620 \pm 0.0006$ is obtained.
Using only 7 CPs results in $r_{7} = 0.2400 \pm 0.0159$.
Inspection of the relative error $100 \cdot (r_i - r_{i - 5}) / r_i$ where $i = 82, 77, 72, ..., 7$ shows that it increases minimally from 0.08 when using 77 CPs to 0.89 when using 32 CPs, but increases substantially to 1.69 when using only 27 CPs, and continues to increase substantially when using fewer CPs.
This indicates that about 30 CPs might be sufficient for registering the EBSD and BSE data, while fewer than 30 CPs would give an incorrect registration and thus an incorrect basis for correlated analysis.
The absolute values for $r$ are different for the two other datasets, but the trends are the same.
As was found by Zhang \textit{et al.} \cite{zhang2014method}, these results show that correction is improved by increasing the number of CPs.
Plots of $r$ and the relative error in $r$ as functions of the number of control points used in image registration are included in the supplementary material.

\subsection{Effects of particles on recovery and recrystallization}

\begin{figure}[htb]
  \centering
  \includegraphics[width=\columnwidth]{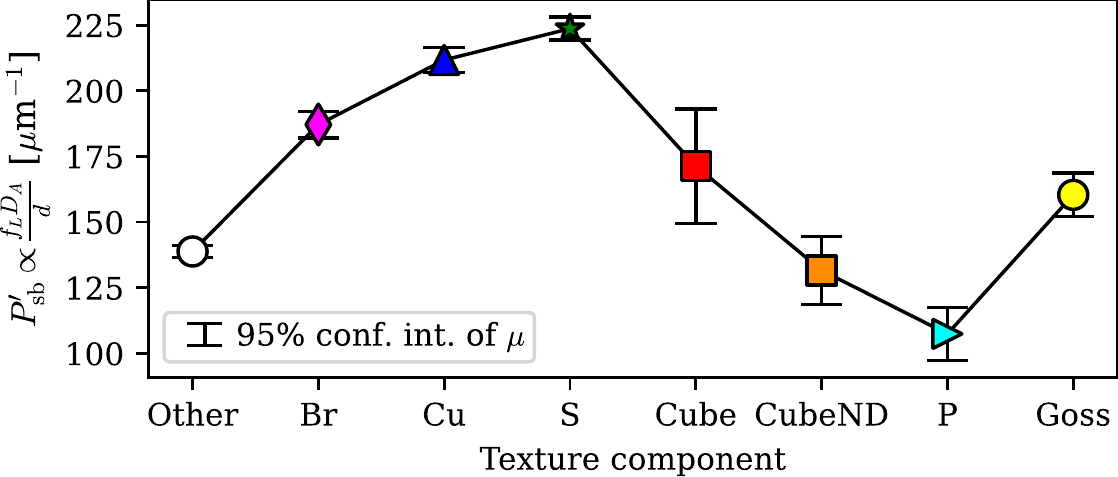}
  \caption{
  Smith-Zener drag at subgrain boundaries per texture component as calculated from Eq. \eqref{eq:smith-zener-drag}.
  }
  \label{fig:smith-zener-drag}
\end{figure}

Recrystallization textures are in general explained by nucleation and growth advantages of grains of certain orientations \cite{doherty1997current}.
The grain data alone shows that Cube subgrains have a size advantage over CubeND and P subgrains since they are respectively 43\% and 33\% larger.
A possible explanation for this observation is that Cube subgrains have experienced faster recovery and started to grow earlier than CubeND and P subgrains \cite{sukhopar2012investigation,bunkholt2019orientation}, however no significant difference in the average $\rho_{\mathrm{GND}}$ between these subgrain populations is observed.
CubeND and P subgrains, on the other hand, have a nucleation advantage, with respectively 214\% and 326\% more subgrains than Cube subgrains.
The combined subgrain and particle data shows that CubeND subgrains have more  dispersoids on their boundaries, while P boundaries are populated by larger dispersoids than the other boundaries of subgrains of recrystallization textures.
The Smith-Zener drag from randomly distributed dispersoids of average radius $d/2$ on near-planar grain boundaries of average energy $\gamma$ is given by $P_{\mathrm{SZ}} = 3\gamma f_{\mathrm{V}}/d$ \cite{rohrer2010introduction,humphreys2017recrystallization}.
The interparticle spacing $N_{\mathrm{s}}^{-0.5}$ of dispersoids in Table \ref{tab:particle-parameters} is about half of the average subgrain size $D_{\mathrm{A}}$, and there is a strong subgrain boundary--dispersoid correlation since 60\% of dispersoids are found at boundaries.
For these reasons, a modified expression for $P_{\mathrm{SZ}}$ must be used \cite{humphreys2017recrystallization}.
Hutchinson and Duggan \cite{hutchinson1978influence} argued that when a large fraction of the particles lies at subgrain boundaries, the drag force per unit area of subgrain boundary (sb) can be rewritten as $P_{\mathrm{sb}} = 3\gamma_{\mathrm{sb}} f_{\mathrm{V}}/(A_{\mathrm{sb}}d^2)$, where $A_{\mathrm{sb}}$ is the area of subgrain boundary per unit volume and $\gamma_{\mathrm{sb}}$ is the average subgrain boundary energy.
If the average subgrain boundary edge length is $L$, we can assume $A_{\mathrm{sb}} = 3/L$ \cite{humphreys2017recrystallization} to get $P_{\mathrm{sb}} = \gamma_{\mathrm{sb}} f_{\mathrm{V}} L/d^2$.
By expressing the dimensionless average dispersoid volume fraction $f_{\mathrm{V}}$ as $f_{\mathrm{L}}d/2$ per texture component, where $f_{\mathrm{L}}$ is the dispersoids per boundary length plotted in Fig. \ref{fig:dispersoids-at-boundaries} (b, c), and by approximating $L$ by the area weighted average subgrain size $D_{\mathrm{A}}$, we arrive at the Smith-Zener drag at subgrain boundaries per texture component

\begin{equation}
  P_{\mathrm{sb}}' = \frac{\gamma_{\mathrm{sb}} f_{\mathrm{L}} D_{\mathrm{A}}}{2d} \propto \frac{f_{\mathrm{L}} D_{\mathrm{A}}}{d},
  \label{eq:smith-zener-drag}
\end{equation}

\noindent where differences in $\gamma_{\mathrm{sb}}$ is neglected.
$P_{\mathrm{sb}}'$ per texture component is plotted in Fig. \ref{fig:smith-zener-drag} with the confidence intervals of $P_{\mathrm{sb}}'$ obtained from error propagation of the confidence intervals of $f_{\mathrm{L}}$, $D_{\mathrm{A}}$, and $d$.
Among subgrains of recrystallization texture, Cube subgrains experience the highest drag while P subgrains experience the lowest drag, with CubeND subgrains in between.
These results help to understand the dominating P texture observed when concurrent precipitation strongly affects the recrystallization texture and grain size, as is the case in the material studied here \cite{tangen2010effect,zhao2016orientation,huang2017controlling}.
It is worth noting that P subgrains have the largest dispersoids on their boundaries, normalized by boundary length, as evident from Fig. \ref{fig:dispersoids-at-boundaries}.

\section{Concluding remarks}

A multimodal data fusion workflow enabling spatially correlated subgrain and particle analysis of alloys in the SEM is presented.
Image registration of BSE images and EBSD maps is accomplished using the thin plate spline transformation function with a sufficient number of control points evenly distributed over the ROI.
Registration and subsequent data fusion results in a distortion corrected EBSD map containing the locations and sizes of particles, many of them smaller than the applied EBSD step size, down to \SI{0.03}{\micro\metre} in diameter.
The workflow is demonstrated on data from a cold-rolled and recovered Al-Mn alloy just before the onset of recrystallization, and results in the following findings:
\begin{enumerate}
  \item P and Cube subgrains experience respectively the lowest and highest Smith-Zener drag from dispersoids on their boundaries among the subgrains of recrystallization texture, which helps to explain the strong P texture in this alloy when non-isothermally annealed or annealed at a low temperature, observed by many.
  \item Subgrains at constituent particles have a growth advantage due to a lower dislocation density, higher boundary misorientation angle $\omega$ and fewer dispersoids on their boundaries.
  \item The dispersoid size per boundary length increases as a function of $\omega$, while the number of dispersoids per boundary length is constant as a function of $\omega$.
\end{enumerate}
The workflow should be applicable to other alloy systems where quantitative correlated analysis of minor secondary phases (of sizes down to and less than \SI{0.1}{\micro\meter}) and orientations of (sub)grains is important to understand microstructure evolution.


\section*{Data and code availability}

The raw EBSD datasets and BSE images required to reproduce these findings are available from Zenodo at \url{https://doi.org/10.5281/zenodo.6470217} [dataset] \cite{anes2022electron}.
All softwares used for analysis of the data, apart from \texttt{MTEX} which requires a \texttt{MATLAB} license, are freely available.
\texttt{Jupyter} notebooks for all processing steps and \texttt{MTEX} scripts can be obtained from \url{https://github.com/hakonanes/correlated-grains-particles-workflow}.


\section*{Acknowledgements}

HWÅ acknowledges NTNU for financial support through the NTNU Aluminium Product Innovation Center (NAPIC).
The authors would like to thank Hydro Aluminium for supplying the material.


\bibliographystyle{elsarticle-num}
\bibliography{library_without_urls}


\end{document}